# Overeducation under different macroeconomic conditions: The case of Spanish university graduates

Maite Blázquez Cuesta    Marco A. Pérez Navarro    Rocío Sánchez-Mangas

Universidad Autónoma de Madrid

## Abstract

This paper examines the incidence and persistence of overeducation in the early careers of Spanish university graduates. We investigate the role played by the business cycle and field of study and their interaction in shaping both phenomena. We also analyse the relevance of specific types of knowledge and skills as driving factors in reducing overeducation risk. We use data from the Survey on the Labour Insertion of University Graduates (EILU) conducted by the Spanish National Statistics Institute in 2014 and 2019. The survey collects rich information on cohorts that graduated in the 2009–2010 and 2014–2015 academic years during the Great Recession and the subsequent economic recovery, respectively. Our results show, first, the relevance of the economic scenario when graduates enter the labour market. Graduation during a recession increased overeducation risk and persistence. Second, a clear heterogeneous pattern occurs across fields of study, with health sciences graduates displaying better performance in terms of both overeducation incidence and persistence and less impact of the business cycle. Third, we find evidence that some transversal skills (language, IT, management) can help to reduce overeducation risk in the absence of specific knowledge required for the job, thus indicating some kind of compensatory role. Finally, our findings have important policy implications. Overeducation, and more importantly overeducation persistence, imply a non-neglectable misallocation of resources. Therefore, policymakers need to address this issue in the design of education and labour market policies.

# 1 Introduction

A highly educated workforce is an essential condition to foster innovation and economic growth: highly educated individuals are more likely to be employed, more productive, earn higher wages, and tend to better cope with economic shocks. Thus, investing in tertiary education (advanced vocational training and university degree programmes) is a relevant tool for promoting economic growth and improving labour market performance. This led the EU to establish the target that at least 40% of the population aged 30–34 should have a tertiary education degree by 2020; a target that was achieved in 2019. However, employability and quality of employment also depend on having specific knowledge and competences that enable individuals not only to access appropriate jobs but to stay in employment and advance in their careers. Hence, an appropriate match between individual's education level, field of education and employment tasks, as well as opportunities for further developing competences are also important to obtain and increase the quality of employment.

Nevertheless, important challenges in this regard still remain in most developed countries. In particular, many tertiary educated graduates lack the skills needed for their successful integration in the labour market, while others end up being mismatched to their jobs (over-educated), especially in the early stages of their careers. Indeed, overeducation has become a widespread phenomenon across Europe, affecting around 22% of the EU-27 workforce in 2019 (including 24.7% of tertiary education graduates aged 25-34 years).[1] Since these education mismatches have potentially large effects on productivity and unemployment, there have been numerous attempts by researchers to identify their main drivers. However, overeducation continues to be a major policy concern, especially during periods of economic recession. In this regard, the 2030 European strategic framework for education and training has highlighted the need to prevent skills gaps and education mismatches. Actions such as those included in the European Skills Agenda have this purpose.

The overarching aim of this paper is to contribute to the recent literature on overeducation by studying how this phenomenon has evolved under different macroeconomic conditions among recent Spanish university graduates. More precisely, we focus on individuals who graduated in 2010 when the Spanish economy was in the midst of the Great Recession that began in 2008, and those who graduated in 2014, when the economy had already entered into a period of expansion. We analyse the incidence and persistence of overeducation under both economic scenarios. In the analysis, we pay special attention to the field of study as a principal factor of overeducation incidence and persistence. Therefore, we study how both indicators are shaped across fields of study and different business cycle conditions.

According to the literature, the field of study is an important determinant of overeducation among tertiary graduates (Caroleo and Pastore 2018). A potential explanation may be that some fields of study involve a higher degree of specialisation than others. For instance, degrees related to scientific and technical fields tend to be more occupationally oriented, which may reduce the likelihood of graduates having to search for jobs outside their own field or that require a lower level of education (Wolbers, 2003; Ortiz and Kucel, 2008). Moreover, the field of study can be considered as an ability signal by employers. Obtaining a degree in scientific and technical fields, which requires a high intellectual capacity, might serve as a signal to employers about applicants' talent and/or motivation (Barone and Ortiz, 2011).

---

[1] Eurostat: https://ec.europa.eu/eurostat/databrowser/view/LFSA_EOQGAC__custom_7629533/default/table?lang=en
Cedefop (European Centre for the Development of Vocational Training):
https://www.cedefop.europa.eu/en/tools/skills-intelligence/over-qualification-rate-tertiary-graduates?country=EU&year=2019#1

An important concern in the overeducation research is whether it is a self-perpetuating state, such that it would be a trap rather than a stepping-stone in the working careers of individuals. In this regard, there is empirical evidence suggesting that overeducation tends to be persistent over time (Rubb, 2003; Frenette, 2004; McGuinness and Wooden, 2009; Baert et al., 2013; among others). Moreover, the field of study seems to condition the transitory or permanent nature of overeducation among tertiary graduates. Some evidence suggests that programmes with a more general orientation exhibit a higher risk of overeducation in the first job compared to more specific programmes, although overeducation seems to be more of a stepping stone in the former case (Verhaest and Van der Velden, 2013). Nevertheless, other studies, such as Albert et al. (2021) or Frenette (2004) for the Spanish and Canadian cases, respectively, obtained lower overeducation persistence among graduates from occupational-specific fields compared to their counterparts from more general fields.

Given the above, overeducation persistence among tertiary-educated people and its relationship with the field of study has become an issue of particular relevance from a policy perspective. If graduates are trapped in mismatched positions, their career prospects might be negatively affected, which may have non-negligible implications for labour market and higher education policies. In such a case, measures should be targeted at assisting these workers towards more efficient educational choices and job search behaviours. Additionally, due to the crowding-out of less educated workers from the labour force by more educated ones, measures aimed at increasing the demand for university graduate labour could also play an important role in reducing overeducation incidence (Albert et al., 2021; Carroll and Tani, 2013; Dolado et al., 2000; Sánchez-Sánchez and Puente, 2020).

Although less explored in the existing literature, the business cycle also seems to play an important role in both the incidence and persistence of overeducation. In particular, economic fluctuations at the time of graduation could be a driving factor behind overeducation among graduates in the early stages of their working careers (Verhaest and Van der Velden, 2013). The research on the impact of the business cycle on overeducation, especially among new labour market entrants, gained attention in the context of the 2008 Great Recession. There is evidence (e.g., Cedefop, 2015) suggesting that the Great Recession led to an increase in the incidence of overeducation among graduates in the UE-28, which rose from 17% in the pre-crisis period (2001–2007) to 28% during the period 2008–2014. Factors related to both supply and demand might explain these figures. From the supply side, it is likely that high unemployment rates during the recession period forced tertiary education graduates to accept jobs that did not match their educational attainments in order to avoid unemployment while searching for jobs that better matched their educational attainments (Wolbers, 2003). Moreover, as the unemployment rate increases so does overeducation, as individuals might decide to invest further in education and training (Pascual et al., 2016). From the demand side, employers who need to cut costs during recessions could prefer to maintain overeducated individuals who could still be useful under better economic conditions (Sánchez-Sánchez and Puente, 2020).

We focus on Spain for several reasons. First, Spain is one of the EU-27 countries with the highest rates of overeducation. Among Spanish tertiary-educated individuals, the incidence of overeducation amounted to 36.1% in 2021, considerably higher than the EU figure of 22.1%.[2] Second, the Spanish case is even more worrisome if we consider the proportion of young tertiary-educated individuals (from 25 to 34 years old). Between 2000 and 2021, the share of people aged 25 to 34 years old with tertiary education in Spain increased by 15 percentage points (pp) from 34% to 49%, a notably higher percentage than the EU average of 42%.[3] We focus our analysis on university graduates that represent approximately 70% of tertiary-educated people in Spain.[4] Finally, the Spanish labour market has shown to be highly dependent on the economic cycle and to a greater extent than other European countries. This was clearly observed during the Great Recession of 2008–2013 and the subsequent recovery period. The unemployment rate in Spain increased from 8.2% in 2007 to 26.1% in 2013, while in the EU-27 it rose from 7.5% in 2007 to 11.4% in 2013 (Eurostat). After several years of economic expansion, in 2019 the unemployment rate reached 14.1% in Spain and 6.7% in the EU-27 (Eurostat).[5]

To the best of our knowledge, research on the relationship between overeducation risk and the economic cycle, the role of field of study, and the extent to which this role is conditioned by economic circumstances is almost non-existent. For Spain, there are a few exceptions in the literature, among them Turmo-Garuz et al. (2019), Acosta-Ballesteros et al. (2018a) or Albert et al. (2021), who partially address these issues. Turmo-Garuz et. al (2019), for instance, studied the incidence of overeducation and its relationship with the business cycle and explain how it differs across fields of study. However, they did not examine the persistence of overeducation across fields and their analysis only comprised university graduates in the region of Catalonia. In contrast, Acosta-Ballesteros et al. (2018a) examined the persistence of overeducation across fields but did not take into consideration the potential impact of the economic cycle and their analysis focuses on all young Spanish workers aged 16 to 34 years old. Finally, Albert et al. (2021) analysed both vertical and horizontal mismatches among Spanish university graduates and the role of field of study in both the incidence and persistence of these job-education mismatches. However, they did not explore the role of the business cycle in shaping the incidence and persistence of overeducation across fields of study. Despite studying underemployment, defined as working fewer hours than desired, and not overeducation, we should also mention the work of Acosta-Ballesteros et al. (2018b), insofar as they examine the importance of both the field of study and the business cycle on labour market performance of young people in Spain at the first stage of their career. Nevertheless, they focus on young workers (aged 16-34) irrespective of their educational attainment.

---

[2] See the CYD Foundation report 2021/2022.
[3] Eurostat: https://ec.europa.eu/eurostat/databrowser/view/SDG_04_20/default/table?lang=en
[4] The tertiary-educated group is composed of university and advanced vocational training graduates. In 2021, the former represented 34% of the population aged 25-34 years old, while the latter represented 14.7% (see Ministerio de Educación y Formación Profesional:
http://estadisticas.mecd.gob.es/EducaJaxiPx/Tabla.htm?path=/laborales/epa/nivfor//l0/&file=nivfor_4_02.px&type=pcaxis&L=0 )
[5] Eurostat: https://ec.europa.eu/eurostat/databrowser/view/UNE_RT_A_H__custom_7010952/default/table?lang=en

To achieve the main purposes of this paper, we use microdata from the Labour Insertion Survey of Recent University Graduates (EILU) in Spain conducted by the National Statistics Institute (INE) in 2014 and 2019. The survey contains rich information on individual characteristics, as well as education and job-related variables. This two-wave data structure allows us to analyse overeducation in both a recession and a recovery period.[6] Moreover, we can identify, at the individual level, job-education mismatches in the first job after graduation and also five years later, which allows us to study overeducation persistence. The role of field of study in both the incidence and persistence of overeducation, as well as its heterogeneous impact in different economic scenarios is also examined. From the methodological point of view, we consider models that explain overeducation risk handling with potential endogeneity issues stemming from various sources: endogenous regressors, individual unobserved heterogeneity and sample selection into employment. To the best of our knowledge, no previous work on this subject has jointly considered these three sources of endogeneity. Finally, another important contribution of our paper is that we examine the role of some transversal skills (language, IT, management and social skills) on the risk of overeducation across fields of study and conditional on the job-education match achieved in terms of theoretical and practical knowledge.

Our research shows that overeducation risk among university graduates in the Spanish scenario is clearly dependent on the economic conditions at the time of graduation. Individuals who graduated in the recession period suffered not only a 10.2 pp higher risk of overeducation than their counterparts who graduated in a recovery period, but also a more significant overeducation persistence five years after graduation (30.2 pp and 12.9 pp, respectively). Moreover, the probability of suffering overeducation is undoubtedly heterogeneous across fields. We find that health sciences graduates are at the lowest risk of overeducation, while arts and humanities graduates exhibit the highest. The differences between health sciences graduates and those from other fields is more pronounced among those who graduated during the recession period, since the former seem to be less dependent on the economic conditions to obtain a good job-education match. A similar conclusion is obtained when looking at overeducation persistence across fields. Finally, our results provide evidence of a certain compensation effect between knowledge and skills to reduce overeducation risk, but with some heterogeneity across fields of study.

This paper is organised as follows. Section 2 reviews the most relevant literature on the topic at hand. Section 3 presents the data and methods used. Section 4 reports and discusses the results. Finally, Section 5 concludes.

---

[6] According to the Spanish Economic Association (2020), the Great Recession in Spain occurred from 2008Q2 to 2013Q2, excluding 2010. We acknowledge that, strictly speaking, the technical conditions to date a recession did not hold in 2010, but the Spanish GDP remained stagnant in 2010, a transition period between two technical recessions. For the sake of simplicity, throughout this paper we will refer to this period as a recession.

## 2   Literature review

Since the pioneering works of Freeman (1976) and Thurow (1975) in the context of the United States, job-education mismatches have been widely studied in the labour and education economics literature, especially for developed countries. A wide number of contributions to this literature have analysed the determinants of vertical mismatches (mainly overeducation) and provided evidence of the impact of various individual, socio-demographic and job-related variables. The works of McGuinness (2006), García-Montalvo (1995, 2009), Leuven and Oosterbeek (2011), Clark et al. (2017), Capsada-Munsech (2017) and Nieto and Ramos (2017), among others, provide some excellent reviews of the extensive literature on overeducation and its determinants.

There is a perception that overeducation predominantly affects university graduates and the existing literature mainly focuses on this direction (see, e.g., Baert, et al., 2013; Carroll and Tani, 2015; Chevalier and Lindley, 2009; Croce and Ghignoni, 2012; and Li and Miller, 2015). From this existing literature, a general pattern can be inferred regarding the factors that increase or diminish the risk of overeducation among university graduates. Among these factors, the field of study plays a key role. Some papers in the literature that have studied the relationship between overeducation and field of study are, among others, Barone and Ortiz (2011), Caroleo and Pastore (2018), Capsada-Munsech (2015), Ortiz and Kucel (2008) or Verhaest et al. (2017). Overall, the evidence on the relationship between overeducation and field of study is mixed and conditioned by the productive structure of the economies. Nonetheless, a common finding in the literature (albeit with a considerable degree of heterogeneity in the methodologies and data used) is that graduates in fields of study with a more general orientation are more likely to fall into overeducation than those in occupation-specific fields (Barone and Ortiz, 2011; Ortiz and Kucel, 2008; Capsada-Munsech, 2017; Rossen et al., 2019). Thus, the evidence suggests that the degree of job-specificity of study programmes significantly shapes the across-field differentials in overeducation risks.

One of the most important issues addressed in the specialised literature is the persistence of overeducation. For many decades, authors have attempted to provide a theoretical background for the study of persistence and justify whether it should be considered a long-lasting or short-term phenomenon. The empirical evidence is mixed in this regard (see Quintini, 2011 for an extensive review). Some papers have found that overeducation is just a temporary phenomenon that most workers overcome through job mobility (Frei and Sousa-Poza, 2012), while others have argued that the phenomenon is quite stable with relatively unlikely successful transitions from overeducation to matched job (see, among others, Rubb, 2003, for the United States; Blázquez and Budría, 2012, for Germany; Kiersztyn, 2013, for Poland). Focusing on university graduates, a vast amount of research has addressed this issue directly or indirectly (see Dolton and Vignoles, 2000; Erdsiek, 2016, 2021; Battu et al., 1999; Verhaest and Omey, 2010; Meroni and Vera-Toscano, 2017; Frenette, 2004; or Verhaest and Van der Velden, 2013). Most of this literature suggests that overeducation among graduates should be considered a persistent problem. For instance, Verhaest and Van der Velden (2013) analysed overeducation in the first five years of the career cycle of college graduates in 13 European countries and Japan and found that among graduates who were overeducated in their job six months after graduation in 2000, 43.3% remained overeducated five years later. However, this figure differs across countries, reaching the highest values among graduates in Japan (66.4%), Switzerland (58.0%) or Germany (53.8%) where overeducation seems to be clearly persistent. In contrast, it could be considered a more temporary problem in France (32.9%), the Czech Republic (32.6%) or the Netherlands (30.2%). In a similar vein, Meroni and Vera-Toscano (2017) used 2005 REFLEX data to investigate overeducation among higher education graduates of 14 European countries who completed their degrees in the 1999–2000 academic year. The authors measured overeducation five years after graduation and found that it is a trap or persistent issue in Southern and Eastern countries, while for Scandinavian and Continental countries the results are heterogeneous. Similarly, Frenette (2004) and Baert et al. (2013) found evidence of the long-lasting nature of overeducation among Canadian and Belgian university graduates, respectively. Clark et. al. (2017) for the US, Savic et al. (2019) for the UK case or Wen and Maani (2019) for the Australian labour market are some of the most recent evidence of the non-transitory nature of overeducation.

Regarding the relationship between overeducation and the economic cycle, the meta-analysis of Groot and Van Den Brink (2000) concluded that the incidence of overeducation is especially related to structural economic fluctuations, such as changes in the labour force growth rate. These economic fluctuations seem to be especially relevant in determining the incidence of overeducation among university graduates at the early stages of their working careers and also have important lasting effects (Summerfield and Theodossiou, 2017). The literature in this regard has mainly followed a cross-country perspective. For instance, Croce and Ghignoni (2012) provided evidence for the period 1998–2006 that cyclical conditions matter for overeducation risk among European university graduates. In particular, they found that the percentage of overeducated graduates reacts significantly to cyclical movements in GDP, such that the risk of overeducation increases in economic downturns. This is in line with the findings of Koppera (2016) for the US for the period 2006–2013 and those of Devereux (2002) for former recessions in the country. However, some exceptions show opposite findings, that is, a higher incidence of overeducation in economic recovery periods than in recessions (see Kiersztyn, 2013 for the Polish case). Despite these exceptions, the overall findings in research examining the impact of the business cycle on overeducation point to a higher risk of overeducation during economic downturns (see the recent works of Pineda-Herrero et al., 2016; Ermini et al., 2017; and Summerfield and Theodossiou, 2017).

The business cycle plays a role not only in the incidence of overeducation, but also in the probability of escaping from it. As Verhaest and van der Velden (2013) showed using a sample of university graduates in 13 European countries and Japan who received their degrees in the 1999–2000 academic year, the stance of the business cycle at the time of graduation influences the extent to which overeducated workers managed to make a transition towards a good match five years after graduation.

## 2.1 Overeducation in the Spanish labour market

The empirical research on overeducation focusing on the Spanish labour market stems from the work of Alba-Ramírez (1993), who concluded that the profile of overeducated workers corresponds to young and highly educated individuals with little job experience. More recent studies on university graduates, such as Albert and Davia (2018) or Albert et al. (2021), have obtained interesting results regarding some specific aspects of overeducation. Regarding gender issues, for example, Albert and Davia (2018) found no evidence of gender differences in the probability of being overeducated in the first job after graduation, while Albert et. al (2021) obtained that women have a higher probability of overeducation in both the first and current job. Concerning other individual characteristics, Albert et al. (2021) showed that having good IT or English skills, having studied abroad or receiving excellence or collaboration grants reduce the risk of overeducation. Additional recent studies, such as Acosta-Ballesteros et al. (2018a) for Spanish individuals younger than 30 years old, including non-graduates, have obtained similar results.

Concerning overeducation persistence in the Spanish labour market, the evidence is mixed. Some papers have found evidence of the transitory nature of overeducation (Alba-Ramírez, 1993; Alba-Ramírez and Blázquez, 2004), while more recent works stress that overeducation is a permanent phenomenon (Acosta-Ballesteros et al., 2018a; Congregado et al., 2016; García-Montalvo, 2013; Rivera Garrido, 2019; Sánchez-Sánchez and Puente, 2020; Albert et al., 2021; and Ramos, 2017). For instance, Albert et al. (2021) showed that, although job mobility allows university graduates to escape from the job-education mismatch, there is still a relevant persistence of overeducation in the early career of those overeducated in the first job.

As regards the impact of the business cycle on the incidence and duration of overeducation, the literature on Spain is scarce. One exception is the work of Sánchez-Sánchez and Puente (2020). The authors compared the phenomenon in an expansion period (2006–2009) and the subsequent recession period (2010–2013) and found that overeducation incidence and persistence were higher during the recession.

Finally, concerning the role of field of study in determining the risk of overeducation among Spanish university graduates, overall, the research confirms that occupational-specific fields (health science; science; and engineering and architecture) exhibit a lower risk (Ortiz and Kucel, 2008; Albert et al., 2021) compared to more generally oriented fields (arts and humanities; and social and legal sciences). This result is in line with other previous works for the case of European college graduates (Verhaest and Van der Velden, 2013). As regards overeducation persistence and its relationship with field of study, the recent works of Albert et al. (2021) and Acosta-Ballesteros (2018a) found a lower persistence of overeducation among Spanish graduates in occupational-specific fields. This evidence is in line with the findings of Frenette (2004) for recent Canadian post-secondary graduates but contradicts the previous results of Verhaest and Van der Velden (2013), who found overeducation among European graduates in less occupational-specific fields to be a stepping stone in their early working careers.

## 3    Data and methods
### 3.1    Dataset and descriptive statistics

The empirical analysis of this study is based on the Spanish Survey on the Labour Insertion of University Graduates (EILU) conducted by the National Statistics Institute (INE). We have used the two available waves of this survey, 2014 and 2019. The first wave includes a sample of 30,379 university graduates who finished their bachelor's degree in the 2009–2010 academic year, while the second one comprises 31,651 individuals who graduated in 2013–2014. In both cases individuals were interviewed around five years after graduation.[7]

This survey has several advantages. First, it contains information on overeducation both in the first job as well as in the job occupied five years after graduation, thus permitting us to analyse overeducation persistence. Second, as mentioned above, it contains information for 2014 and 2019 about those who graduated in 2010 and 2014, respectively, which allows us to study differences in the incidence and persistence of overeducation among tertiary education graduates under different macroeconomic conditions. Finally, the dataset contains rich information on the field of study; a factor that has been shown to play a key role in the incidence of overeducation among tertiary graduates (Acosta-Ballesteros et al., 2018a; Barone and Ortiz, 2011; Caroleo and Pastore 2018; Capsada-Munsech, 2015; Ortiz and Kucel, 2008; Verhaest et al., 2017). We analyse the role of field of study on overeducation incidence and persistence, as well as to what extent these phenomena are shaped by the business cycle.

Our measure of overeducation relies on self-assessed information provided by workers. In particular, individuals are considered to be overeducated based on their response regarding the suitability of their educational level to the job-education requirements. We opt for this approach among the three alternatives that have been commonly used in the literature to measure overeducation: the objective/job approach (Eckaus, 1964), the statistical approach (Clogg and Shockey 1984; Verdugo and Verdugo, 1989) and the subjective approach (Duncan and Hofman, 1981). To date, no consensus has been reached as to which is the best approach to measure overeducation, since each one has been shown to have its own advantages and drawbacks (see Flisi et al., 2017; and Table 1 in Capsada-Munsech, 2019 for a summary). In this paper we follow key studies in the literature of overeducation that have adopted the self-assessment/subjective approach (Duncan and Hoffman, 1981; Hartog and Oosterbeek, 1988; Sicherman, 1991; Battu et al., 1999; Battu et al., 2000; Allen and Van der Velden, 2001; Green and Zhu, 2010; Frei and Sousa-Poza, 2012; and Baert et al., 2013; among others).

Several additional aspects are also worth noting. First, our two-wave data correspond to graduates who obtained their degree before and after, respectively, the Bologna Process of cross-country harmonisation of some characteristics of higher education degree programmes in European countries. Thus, our results might inevitably be affected by this issue. However, several study-related control variables have been considered to minimise the potential problems stemming from this approach. Among others, we use an indicator of whether a student has done internships as part of their degree programme, which is one of the most relevant measures aimed by the Bologna Process. However, the proportion of individuals who did an internship as part of their degree programme is quite similar in both waves and even higher among 2010 graduates (64.21% for 2010 graduates and 60.51% for 2014 graduates). Indeed, there is empirical evidence suggesting that the education obtained under the Bologna Process does not enhance the employability of

---

[7] More specifically, interviews were conducted between four and a half and five and a half years after graduation depending on the wave. For the sake of simplicity, we refer to five years after graduation throughout the paper.

Spanish university graduates compared to the pre-Bologna higher education degree programmes (Canal Domínguez and Rodríguez Gutiérrez, 2022). Consequently, working with data on both types of graduates (pre- and post-Bologna Process) does not seem to be a major issue in our analysis. A second issue to keep in mind is that we do not have precise details on when the first job was found. However, the data show that 73.7% of the individuals who have ever had a paid job in the period of analysis found their first job during the first year after graduation. Third, it is also worth mentioning that almost all the interviewed individuals (nearly 95.5% of the sample) have ever had a paid job, but only around 79.0% was working at the time of the interview. The latter figure is quite different for individuals interviewed in 2014 (73.1%) and in 2019 (84.8%), thus reflecting the importance of macroeconomic conditions. Therefore, potential bias due to sample selection into employment must be taken into account when studying overeducation at the time of the interview.

Table 1 describes the variables included in our analysis. The variables are grouped into different categories and the definition and values provided for each one.

**Table 1. Variables definition**

| A. Job-education mismatches – related variables | |
|---|---|
| *Variable* | *Definition* |
| Overeducation in the first job | Self-assessed overeducation in the first job after graduation: the worker reports that the level of education required for their job is lower than a bachelor's degree. Binary indicator (1 = yes, 0 = no). |
| Overeducation in the current job | Self-assessed overeducation in the job at the time of the interview. Binary indicator (1 = yes, 0 = no). |
| **B. Individual characteristics** | |
| *Variable* | *Definition* |
| Male | 1 if male, 0 otherwise. |
| Age | Different age groups: 1 (<30 years old), 2 (30–34 years old), 3 (≥35 years old). |
| Spanish | 1 if Spanish, 0 otherwise. |
| ICT | ICT knowledge: 1 (Basic), 2 (Advanced), 3 (Expert). |
| Languages spoken | 1 if the individual speaks two or more languages, 0 otherwise. |
| **C. Study-related variables** | |
| *Variable* | *Definition* |
| Studied abroad | 1 if the graduate has studied abroad, 0 otherwise. |
| Collaboration or excellence grant | 1 if the graduate has obtained an excellence or collaboration grant during the degree, 0 otherwise. |
| Private university | 1 if university of origin is a private university, 0 if public. |
| Field of study | Field of study: 1 (Arts and humanities), 2 (Social and legal sciences), 3 (Science), 4 (Engineering and architecture), 5 (Health sciences). |
| Internship outside the degree programme | 1 if the graduate has done an internship outside the degree plan, 0 otherwise. |
| Postgraduate degree | 1 if the graduate has a postgraduate degree, 0 otherwise. |

**Table 1. Variables definition (cont.)**

**D. Job-related variables**

| Variable | Definition |
|---|---|
| Part-time | 1 if the graduate works part-time, 0 otherwise. |
| Professional situation | Professional situation in the individual's job. Type of contract: 1 (Trainee), 2 (Permanent contract), 3 (Fixed contract) |
| Experience | 1 if the graduate has 2 or more years of professional experience at the time of the interview, 0 otherwise. |
| Employers | 1 if the graduate has had 2 or more employers at the time of the interview, 0 otherwise. |
| Theoretical knowledge | Relevance of theoretical knowledge to obtain the current job: 1 (Not important), 2 (Slightly important), 3 (Moderately important), 4 (Important), 5 (Very important). |
| Practical knowledge | Relevance of practical knowledge to obtain the current job: 1 (Not important), 2 (Slightly important), 3 (Moderately important), 4 (Important), 5 (Very important). |
| Language skills | Relevance of language skills to obtain the current job: 1 (Not important), 2 (Slightly important), 3 (Moderately important), 4 (Important), 5 (Very important). |
| IT skills | Relevance of IT skills to obtain the current job: 1 (Not important), 2 (Slightly important), 3 (Moderately important), 4 (Important), 5 (Very important). |
| Social skills | Relevance of social skills to obtain the current job: 1 (Not important), 2 (Slightly important), 3 (Moderately important), 4 (Important), 5 (Very important). |
| Management skills | Relevance of management skills to obtain the current job: 1 (Not important), 2 (Slightly important), 3 (Moderately important), 4 (Important), 5 (Very important). |

After eliminating individuals with military occupations, independent workers, individuals who work helping in a family business, those whose current and/or first job is outside the European Union (or the United Kingdom) and observations with missing values in the variables of interest, we end up with a sample of 37,819 graduates employed at the time of the interview.[8] Table 2 and Table A1 in the Appendix provide the descriptive statistics of the main variables used in the analysis for this sample of individuals. We report the mean and standard deviation of the variables for the pooled sample.

---

[8] The sample cleaning led us to discard 24.1% of the graduates employed at the time of the interview from the original dataset.

**Table 2. Summary of descriptive statistics**

| A. Job-education mismatches – related variables | | | C. Study-related variables | | |
|---|---|---|---|---|---|
| *Variable* | *Mean* | *SD* | *Variable* | *Mean* | *SD* |
| Overeducation in the first job | 0.284 | 0.451 | Studied abroad | 0.158 | 0.365 |
| Overeducation in the current job | 0.169 | 0.375 | Excellence or collab. grant | 0.076 | 0.265 |
| | | | Private university | 0.145 | 0.352 |
| **B. Individual characteristics** | | | Field of study | | |
| *Variable* | *Mean* | *SD* |   Arts and humanities | 0.084 | 0.280 |
| Male | 0.410 | 0.492 |   Science | 0.093 | 0.290 |
| Age | | |   Social and legal sciences | 0.444 | 0.497 |
|   <30 years old | 0.549 | 0.498 |   Engineering and architecture | 0.232 | 0.422 |
|   30–34 years old | 0.264 | 0.441 |   Health sciences | 0.147 | 0.354 |
|   ≥35 years old | 0.186 | 0.390 | Internship outside degree | 0.299 | 0.458 |
| Spanish | 0.992 | 0.091 | Postgraduate degree | 0.427 | 0.495 |
| ICT knowledge | | | | | |
|   Basic | 0.135 | 0.342 | **D. Job-related variables**[a] | | |
|   Advanced | 0.661 | 0.473 | *Variable* | *Mean* | *SD* |
|   Expert | 0.204 | 0.403 | ≥2 years of experience | 0.858 | 0.349 |
| Two or more languages spoken | 0.943 | 0.231 | ≥2 employers | 0.709 | 0.454 |
| | | | Type of journey in current job | 0.182 | 0.386 |
| | | | Professional situation in current job | | |
| | | |   Trainee | 0.100 | 0.300 |
| | | |   Permanent contract | 0.560 | 0.496 |
| | | |   Fixed contract | 0.341 | 0.474 |

*Notes*: N = 37,819 observations.
[a] For the sake of simplicity, descriptive statistics for the knowledge and skills variables are reported in Table A1 of the Appendix.

According to the figures in Table 2 and in Table A1 of the Appendix, 99.2% of the individuals are Spanish and 59.0% are women. Regarding age, 54.9% were younger than 30 years old at the time of the interview, 26.4% were aged between 30 and 34 years old and 18.6% were above the age of 34. Only 29.9% of the individuals did an internship outside their degree programme, 14.5% obtained their degree at a private university, 15.8% studied abroad and 7.6% was awarded an excellence or collaboration grant. Regarding the field of study, 8.4% of university graduates studied arts and humanities and a quite similar proportion studied science (9.3%), while social and legal sciences was the field chosen by 44.4% of individuals. Finally, 23.2% of the sample comprises individuals who studied engineering and architecture and the remaining 14.7% studied health sciences.[9]

Focusing on the variables concerning overeducation, around 28.4% of the individuals reported being overeducated in the first job. However, for overeducation in the current job, this figure drops to 16.9%. Figure 1 offers descriptive evidence of the incidence of overeducation for graduates across fields of study and for each wave. We report the incidence of overeducation both in the first job after graduation and in the job at the time of the interview, thus allowing a first descriptive insight about the heterogeneity in overeducation persistence, across fields of study and under different macroeconomic conditions.

---

[9] The presence of women across fields shows some interesting patterns. It is remarkable that the proportion of women in the fields of social sciences (67%) and health sciences (75%) is much higher than the proportion of women in the whole sample, while the opposite occurs in the fields of engineering and architecture where the proportion of women is only 31%.

**Figure 1: Incidence of overeducation across fields of study**

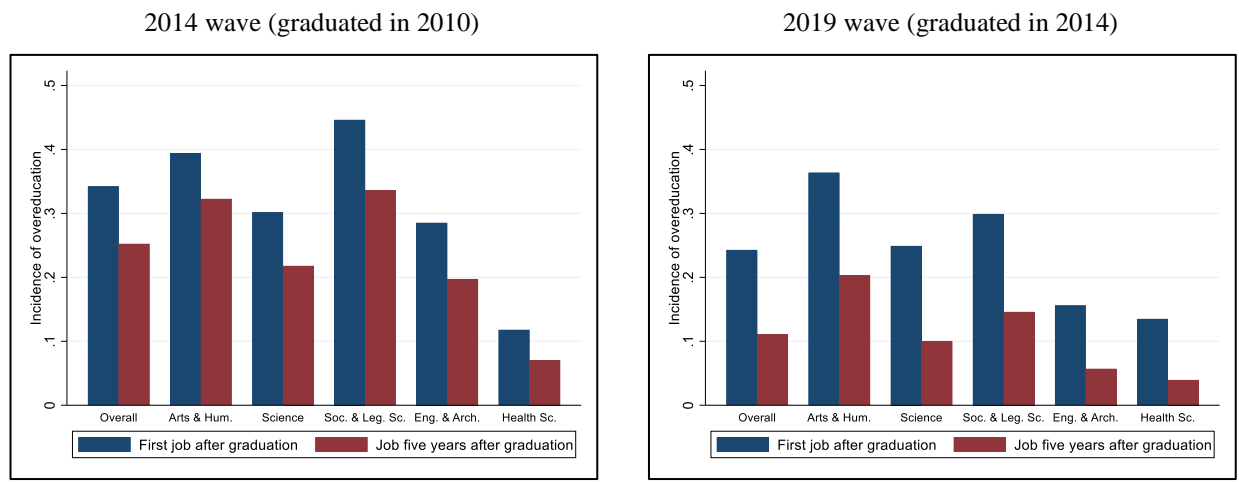

The descriptive analysis reveals that the incidence of overeducation among recent graduates was higher when the current economic conditions were worse. As can be seen, 34.24% of the individuals who graduated in 2010 and were employed at the time of the interview reported being overeducated in their first job. However, the corresponding figure for those who graduated in 2014 was only 24.28%. Moreover, these differences in the incidence rate associated with the economic situation at the time of graduation persisted five years after graduation, with overeducation rates of 25.23% and 11.04% for individuals who obtained their degree in 2010 and 2014, respectively. Moreover, the reduction in the incidence of overeducation between the first and current job is more pronounced for the 2014 graduates than for those who graduated in 2010 (13.24 pp vs. 9.01 pp). Thus, overeducation seems to be more persistent for those who graduated during a recession period. This might be explained by the fact that the more favourable macroeconomic conditions experienced by graduates in 2014 could have allowed them to get not only better matching opportunities in the labour market, but also better chances to improve that matching in the early career stage compared to their counterparts in a worse economic scenario.

Concerning fields of study, graduates from occupation-specific fields such as health sciences have the lowest incidence of overeducation in the first and in the current job, regardless of the period analysed. In contrast, graduates in arts and humanities and social and legal sciences display the highest incidence. This is in line with the evidence described in the literature, where occupation-specific fields have been found to exhibit a lower risk of overeducation than fields with a more general orientation (Ortiz and Kucel, 2008; Verhaest and Van der Velden, 2013; Albert et al., 2021). We also observe a clear difference between waves. Individuals who graduated in 2014 have a notably lower incidence of overeducation five years after graduation than those who graduated in 2010, with more pronounced differences found among graduates in the fields of social and legal sciences, engineering and architecture. The higher risk of overeducation suffered by 2010 graduates in engineering and architecture compared to their counterparts who graduated in 2014, could be partially explained by the huge job destruction rate in the Spanish construction sector around the time of the Great Recession. This sector demanded a significant number of engineering and architecture graduates (Turmo-Garuz et al., 2019) who subsequently experienced a higher risk of overeducation once the construction sector collapsed.

Overall, the differences in the risk of overeducation across waves are not so prominent at labour market entry as they are five years after graduation, which could suggest the more long-lasting nature of overeducation during recession periods. For instance, arts and humanities graduates show a very similar figure for overeducation in the first job in both waves (39% and 36%), but their overeducation incidence five years after graduation is markedly lower for those who graduated under better economic conditions.

## 3.2 Methods and model specification

Our outcome of interest is overeducation at the time of the interview and the population of interest is composed of young graduates. We want to explore the role that overeducation in the first job, the field of study and the economic cycle play in this phenomenon. Two important econometric concerns deserve attention in this analysis. The first concern is the selection of individuals into employment, since, as we mentioned, workers at the time of the interview accounted for around 79% of the sample. The second concern is related to potential endogeneity, since overeducation in the first job might be correlated with unobservable factors, such as personality traits or ability, which can determine both the probability of having a job at the time of the interview and the probability of suffering overeducation in that job.

To account for these two potential sources of bias, we consider a system of three equations, which is estimated simultaneously as a trivariate probit model given by:

$$y_c = 1(\alpha y_f + \beta' x_c + u_c \geq 0)$$

$$E = 1(\delta y_f + \gamma' x_E + u_E \geq 0)$$

$$y_f = 1(\lambda' x_f + u_f \geq 0)$$

where ($y_c$, $E$, $y_f$) are the observed binary indicators for overeducation in the current job (at the time of the interview), employment status at that time and overeducation in the first job, respectively. Our model is a recursive trivariate probit model, where $y_f$ is a potential endogenous regressor not only in the equation of interest for $y_c$ but also in the selection equation into employment at the time of the interview.

The equation for $y_c$ includes the potential endogenous regressor of overeducation in the first job ($y_f$), which allows us to capture overeducation persistence. A 2019 wave indicator (recovery period) is also included to control for the role of the business cycle. Additionally, we include binary indicators for each field of study and a set of controls denoted as $x_c$ that comprises individual characteristics (gender, age, ICT knowledge, languages, Spanish nationality), study-related variables (studies abroad, collaboration or excellence grant, private university, internship outside the degree programme, postgraduate degree) and job-related variables for the job at the time of the interview (region or country of the job, type of contract, part-time job, experience, number of employers, skills variables). We also consider some interaction terms. First, we have included an interaction between the dummy variables that capture the recovery period and fields of study to allow for potential differences in the role of the field of study on the probability of overeducation under different macroeconomic conditions. Second, we have included an interaction between the dummy variables that capture the recovery period and being overeducated in the first job to analyse overeducation persistence in the different economic periods. Finally, we have added an interaction term between the field of study and overeducation in the first job to capture potential differences in overeducation persistence across fields of study.

Vector $x_E$ for the equation that captures selection into employment at the time of the interview includes individual characteristics, study-related variables, the recovery indicator, the interaction of this latter with the field of study and, finally, the unemployment rate among young tertiary graduates of the region of residence. We also include an interaction term between $y_f$ and the other two variables of interest (field of study and the recovery indicator).

Vector $x_f$ for the first job overeducation equation contains similar variables to those in $x_c$ for the current job but refers to the first job and has some differences. For instance, we cannot incorporate information in $x_f$ concerning the job-related skills variables, experience or number of employers, since these variables are not available for the first job. In contrast, we include a set of variables in this vector that describe the job-search method. Finally, we have included interaction terms between the economic period and fields of study.

The error terms of the three equations ($u_c$, $u_E$, $u_f$) are assumed to follow a trivariate normal distribution with zero-mean, unit variances, and correlation coefficients potentially different from zero. We estimate the model by simulated maximum likelihood using the GHK algorithm (Geweke 1989; Hajivassiliou and McFadden 1998; and Keane 1994).

Although the nonlinearity of our model guarantees identification without imposing exclusion restrictions (Wilde, 2000; Greene and Zhang, 2019), in line with other authors we consider additional covariates in both the equation for overeducation in the first job and the selection equation (see Wooldridge, 2010; Mourifié and Meango, 2014; and Han and Vytlacil, 2017 for a similar strategy in different contexts). As regards the equation for overeducation in the first job, we have included some job-related characteristics only referred to the first job. Specifically, the job-search method, the time needed to find the first job after graduation, the Spanish region or foreign country of the first job, the type of contract and part-time or full-time work regime. Concerning the equation for selection into employment, we have included the unemployment rate for young tertiary educated individuals of the region of residence (or country for those living overseas) as an exclusion restriction. Instruments of this type have been used in part of the literature that analyses labour outcomes accounting for endogeneity problems (Triventi, 2014; Rubb, 2014).

# 4 Estimation results

In this section we present the estimation results of the recursive trivariate probit model for the probability of suffering overeducation in the graduates' current job, controlling for the potential endogeneity of overeducation in the first job as well as for sample selection issues.

Table 3 shows the estimated average partial effects of the variables of interest, i.e. those capturing both the impact of the business cycle and the persistence of overeducation across fields of study.[10] It is important to notice that, although our estimates are obtained once controlling for a wide set of observable factors and taking into account potential sources of endogeneity, they should be interpreted with some caution insofar as we cannot preclude the possibility of other unobserved factors affecting our results, which would prevent us from interpreting them as causal effects.

**Table 3. Estimated APE on the probability of overeducation in the job at the time of the interview**

|  | Full sample | 2014 wave (Graduated in 2010) | 2019 wave (Graduated in 2014) |
|---|---|---|---|
| Recovery period | -0.106***(0.018) |  |  |
| Overeducation in the first job | 0.201***(0.013) | 0.302***(0.020) | 0.129***(0.012) |
| Field of study (Ref: Arts and humanities) |  |  |  |
|    Science | -0.035***(0.005) | -0.021***(0.005) | -0.046***(0.008) |
|    Social and legal sciences | -0.019***(0.005) | -0.006(0.007) | -0.030***(0.006) |
|    Engineering and architecture | -0.060***(0.008) | -0.042***(0.011) | -0.076***(0.012) |
|    Health sciences | -0.109***(0.010) | -0.124***(0.015) | -0.098***(0.010) |
| $corr(u_f, u_E)$ | 0.354***(0.057) |  |  |
| $corr(u_f, u_c)$ | 0.283***(0.024) |  |  |
| $corr(u_E, u_c)$ | -0.066(0.050) |  |  |
| Observations | 37819 | 37819 | 37819 |
| Wald test, $p$-value | 0.000 | 0.000 | 0.000 |

*Notes:* APE = average partial effects from a trivariate probit model estimation. Clustered standard errors by occupation are obtained by the Delta method and shown in parentheses. Individual characteristics, study-related variables, job-related variables, interaction terms (see Section 3.2) and regional dummies are included in all models. The full set of results are reported in Table A2 of the Appendix.
\*\*\*, \*\*, \*: statistically significant at 1%, 5% and 10%, respectively.

Our estimation results reveal that the business cycle exerts an important effect on the risk of overeducation among Spanish university graduates. In particular, we obtain that, once we control for individual characteristics as well as study-related and job-related variables, those who graduated in 2014 during the recovery period are 10.6 pp less likely to be overeducated at early stages of their professional career than those who graduated in 2010 during the recession period. These results are in consonance with most of the previous literature on this topic for university graduates. For instance, Croce and Ghignoni (2012) analysed a panel of European countries and obtained that economic downturns are associated with a higher incidence of overeducation. In a similar vein, Summerfield and Theodossiou (2017) suggested for the German case that increases in unemployment rates at graduation contribute to higher risks of overeducation in the future. Other works, such as the meta-analysis performed by Verhaest and Van der Velden (2013) for 14 countries, are in the same line.

---
[10] Table A2 of the Appendix shows the estimated average partial effects of other variables considered in the specification.

Concerning the analysis of field of study, we obtain that graduates from occupation-oriented fields suffer a lower overeducation risk than fields with programmes of a more general orientation. This result is in line with previous works in the literature (see Albert and Davia, 2018; Blázquez and Mora, 2010; Ortiz and Kucel, 2008; Capsada-Munsech, 2017; among others). In particular, we find that health sciences graduates exhibit the lowest risk of overeducation five years after graduation. Specifically, this group has a 10.9 pp lower risk than arts and humanities graduates (the reference category). We also obtain that engineering and architecture or sciences graduates have a significantly lower risk of overeducation (6.0 pp and 3.5 pp, respectively) than arts and humanities graduates. For Spanish university graduates, Albert and Davia (2018) show that those in STEM (science, technology, engineering, and mathematics) and health sciences are generally more protected from overeducation than graduates in other fields, such as the social sciences and arts and humanities. Our findings suggest that social and legal sciences graduates seem to benefit more from the better economic performance than their counterparts from the arts and humanities field. It is also remarkable that the difference in the risk of overeducation for health sciences graduates with respect to arts and humanities shrinks in the recovery period, contrary to what occurs in the rest of the fields. This might be explained by the lesser dependence of these fields on macroeconomic conditions. Finally, it is important to note that health sciences and engineering and architecture graduates report the lowest probability of suffering overeducation in both sample periods. For instance, in 2019 engineering and architecture and health sciences graduates showed a 7.6 pp and 9.8 pp lower risk of overeducation, respectively, than arts and humanities graduates. The lower risk of overeducation in technical or scientific fields, such as health sciences or engineering, could be explained by the fact that they are aimed at very specific occupations requiring discipline-specific skills, while arts and humanities or social sciences have a wider scope.

Figure 2 shows the risk of overeducation for graduates in the recovery period with respect to their counterparts in recession for all fields of study. Clearly, the economic circumstances seem to condition the role of the field of study in determining overeducation risk. For instance, those who graduated in engineering and architecture in 2014 show the highest reduction in the risk of overeducation compared to their counterparts who graduated in 2010. This reduction might be partially explained by the collapse of the construction sector in Spain after the Great Recession, as we have already mentioned in previous sections. The findings are only slightly different for those with a degree in arts and humanities, sciences or social and legal sciences. In contrast, the impact of the business cycle on overeducation risk is clearly lower for health sciences graduates. This might be related to the fact that employment in the health sector in Spain has been less affect than other sectors by the negative consequences of the Great Recession as the International Monetary Fund (2017) has suggested.

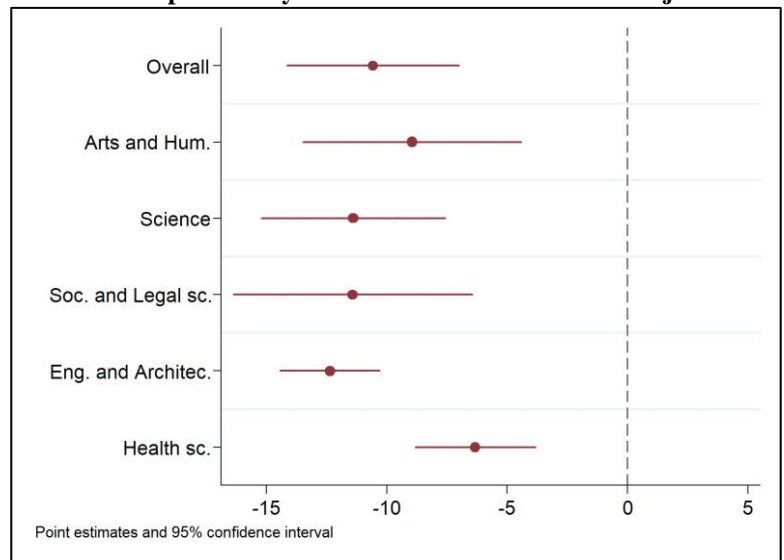

**Figure 2: Impact of macroeconomic conditions on the probability of overeducation in the current job**

*Note*: Average partial effect in percentage points, expansion period vs. recession period.

Overall, our results regarding the impact of macroeconomic conditions on the risk of overeducation provide some evidence of the crowding out effect as posited by the Job Competition Theory (Thurow, 1975), insofar as those who graduated and entered the labour market under a troublesome economic scenario characterised by high unemployment rates exhibited the highest risk of overeducation in the early stages of their professional careers than those who graduated in a recovery period. This result is in line with Pompei and Selezneva (2021) who suggest that overeducation can act as a moderator in recession periods in EU countries, favouring the labour market entry of young individuals with more years of education. These authors find that returns to schooling are higher in countries with high incidence of job-education mismatches. Additionally, our results are in line with other papers that provide evidence of the signalling role of overeducation to enhance individuals' employability and bargaining power (Garcia-Mainar and Montuenga, 2019; Charlot et al., 2005).

As stated throughout this article, persistence is a key element in the overeducation literature. When pooling both sample periods, we obtain that graduates who were overeducated in their first job have a 20.1 pp higher probability of remaining overeducated in their job five years after graduation. This shows that overeducation should be considered as a persistent phenomenon among recent graduates in Spain, which is consistent with most of the literature for Spain (Congregado et al., 2016; García-Montalvo, 2013; Acosta-Ballesteros et al., 2018a; Rivera Garrido, 2019; Sánchez-Sánchez and Puente, 2020; Albert et al., 2021).

Table 3 shows how this persistence of overeducation varies under different macroeconomic conditions. In particular, being overeducated in the first job after graduation increases the probability of remaining overeducated five years after graduation by 30.2 pp for individuals who graduated in 2010 and only by 12.9 pp for those who graduated in 2014. Therefore, these findings suggest that entering the labour market under a troublesome economic scenario has long lasting effects on the probability of achieving a good job-education match. This result is in line with other works on the Spanish labour market (Sánchez-Sánchez and Puente, 2020).

Moreover, as can be observed in Figure 3, the pattern persists across fields of study and is conditioned by the business cycle. No significant differences across fields can be observed for those who graduated during the recession, except for the case of health sciences graduates for whom overeducation seems to be a less persistent issue. In particular, the 2010 health sciences graduates who were overeducated in their first job exhibited a 20.8 pp higher probability of remaining overeducated in their job at the time of the interview compared to the non-overeducated in their first job. This estimated effect is clearly below the effects for the rest of the fields (36.1 pp for arts and humanities, 32.4 pp for social sciences, 31.7 pp for science and 30.5 pp for engineering and architecture). Although the lowest risk of overeducation persistence among health science graduates remains for those who graduated in a recovery scenario (8.2 pp), the difference with other fields of study is of lower magnitude. For instance, the 2014 engineering and architecture graduates who suffered overeducation in their first job increased their probability of overeducation by only 10.5 pp five years later. A higher figure is observed for graduates from other fields, such as social and legal sciences or science, with 14.5 pp and 13.3 pp, respectively. On the other hand, and regardless of the better macroeconomic conditions, arts and humanities graduates still exhibit an important difference with the rest of the fields, while suffering the highest risk of remaining overeducated five years after graduation (19.5 pp). In addition, and in line with other authors (Meroni and Vera-Toscano, 2017; Acosta-Ballesteros et al., 2018a), the findings shown in Figure 3 provide evidence that fields which impart more occupation-oriented knowledge and skills are associated with a lower incidence and persistence of overeducation.

In sum, our results suggest that the long-lasting nature of the overeducation phenomenon in the Spanish labour market needs to be approached jointly considering the field of study and the macroeconomic conditions.

**Figure 3: Impact of being overeducated in the first job on the probability of overeducation in the current job**

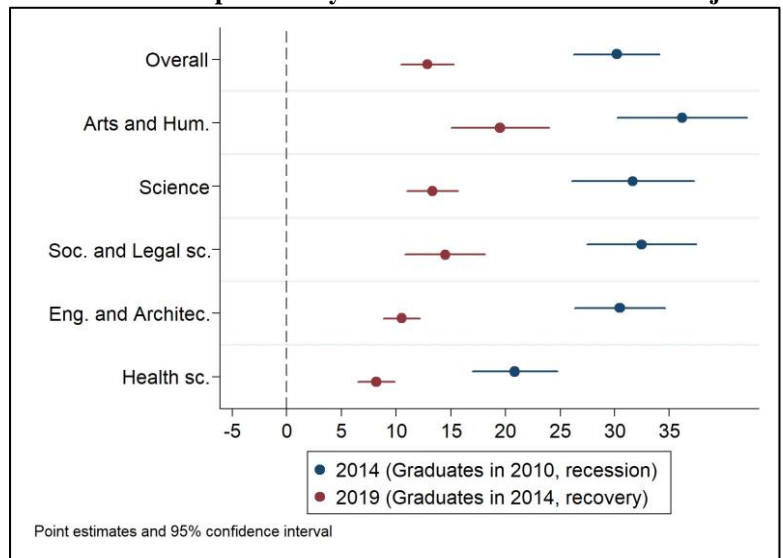

*Note*: Average partial effect in percentage points.

Looking at our results for the rest of overeducation determinants, available in Table A2 of the Appendix, we obtain some interesting findings in line with most of the recent literature. Concerning individual characteristics, we find that being over 34 years old (compared to individuals under 30 years old) is associated with an increase in the probability of suffering overeducation, while having studied abroad in a private university or having postgraduate studies reduce the risk of overeducation. These results are in line with other studies for Spanish university graduates, namely Albert and Davia (2018). Moreover, Capsada-Munsech (2017), and Barone and Ortiz (2011) suggested that the role of postgraduate studies is especially relevant in countries where there is a high percentage of tertiary-educated individuals, such as Spain. We obtain no significant gender differences in the risk of overeducation among Spanish university graduates; a result that is in line with part of the related literature for university graduates (Albert and Davia, 2018; Groleau and Smith, 2019). Finally, it is also worth noticing the role of the job-related variables included in the analysis. In particular, the work schedule, professional situation and current job location play a significant role in the probability of overeducation. We find that working part time (versus full time) is associated with an increase in the risk of overeducation, which is in line with previous evidence in the literature (see Albert et al., 2021; Carroll and Tani, 2015). Trainees have a lower probability of being overeducated than those with a permanent or fixed-term contract. This result seems reasonable since internships are usually closely related to individuals' education. Concerning the job location, our results show that individuals working in Spanish regions with higher unemployment rates or lower job opportunities suffer a higher risk of overeducation.

We have also performed several sensitivity analyses to test the robustness of our findings. First, to account for sex segregation across fields of study (see Zafar, 2013), we include interactions between gender and fields to capture potential gender differences in our main results. Moreover, regarding standard errors, we considered alternative clustering schemes including regions and a more disaggregated classification of occupations. Both strategies leave our results mainly unchanged. Second, given the lack of evidence of a correlation between the error term of the overeducation equation in the current job and the employment equation, we might believe that sample selection is not of major importance in our model. Thus, we estimate a bivariate probit model without considering this source of endogeneity. Not surprisingly, the results are similar to those reported in this section (see Table A3 of the Appendix). The alternative estimation of a bivariate model for both employment and overeducation in the current job leads to interesting findings regarding sample selection. We find evidence of this selection issue only if overeducation in the first job is not included in the equation for overeducation in the current job. This suggests that the inclusion of this regressor, which is key for our analysis of overeducation persistence, seems to capture – at least to some extent – the correlation stemming from the potential selection into employment. However, given the evidence of correlation between unobservable factors affecting overeducation in the first job and current employment status, we opt for the more flexible approach offered by the trivariate model.

## 4.1 Overeducation and career competencies

The non-negligible rate of overeducation risk and its persistence among graduates suggest that an important priority in the transition from higher education to the labour market should be to strengthen graduates' career through engaging them in developing "career competencies" that can aid them in finding adequate employment.

Firms have to continuously adapt to ever changing demands, which requires increasing flexibility of the workforce as well as matching job skills with new requirements. Career competencies, which can be seen as a cluster of related knowledge, skills and attitudes (Akkermans et al., 2013), are likely to play a crucial role in this adaptative process (Arthur and Rousseau, 1996). Indeed, such competencies have been shown to be a critical tool for employability enhancement in higher education (Blokker et al., 2019; Bridgstock, 2009; Clarke, 2018; Okay-Somerville and Scholarios, 2017). Empirical evidence has shown that they facilitate a smooth transition to the labour market (Kuijpers & Meijers, 2012) and contribute to career success (Eby et al., 2003).

Therefore, identifying potential interconnections between both theoretical and practical knowledge and skills that contribute to improving graduates' performance on the job would be of paramount importance for educational and labour market policymakers. This section is intended to partially address this issue. In particular, we examine the potential interconnection of theoretical and practical knowledge – inherent to a specific field of study – and other "transversal skills and competencies" in determining the risk of overeducation. In this respect, in 2019–2020 the skills pillar of the European Skills, Competences, Qualifications and Occupations (ESCO) classification agreed that knowledge concepts (the body of facts, principles, theories and practices that is related to a field of work or study) should be considered separately from "transversal skills and competences". This reflects the fact that these transversal skills, by definition, are independent of any specific field of work, study or activity. However, it is likely that the development of different transversal skills leads to different outcomes in the labour market depending on the specific field of study and the job-education match achieved in terms of theoretical and practical knowledge inherent to a given field.

Thus, this subsection examines whether the effect of certain transversal skills[11] (language, IT, management and social skills) on the risk of overeducation varies depending on the fit achieved between the theoretical and practical knowledge acquired in the different fields of study and required for the job position. The latter is captured by self-reported information that refers to "the relevance of theoretical and practical knowledge to get the current job". We understand that declaring as relevant a specific type of knowledge or skill means that the individual has that knowledge or skill and hence an appropriate match is achieved. Figure 4 summarises the role of theoretical and practical knowledge as well as different types of transversal skills on the probability of overeducation. We present the results across fields of study.[12]

---

[11] Transversal skills refer to 'transferable', 'soft', non-cognitive' and 'socio-emotional' skills. According to the ESCO definition: "Transversal skills and competences (TSCs) are learned and proven abilities which are commonly seen as necessary or valuable for effective action in virtually any kind of work, learning or life activity. They are "transversal" because they are not exclusively related to any particular context (job, occupation, academic discipline, civic or community engagement, occupational sector, group of occupational sectors, etc.)".

Figure 4A shows that, overall, both theoretical and practical knowledge play an important role in reducing the risk of overeducation, with theoretical knowledge being more marked. The behaviour is quite homogenous across fields of study concerning practical knowledge, but some heterogeneity appears when looking at the theoretical side, with reductions in the probability of overeducation that go from around 8 pp for health sciences graduates to 20pp for arts and humanities graduates when this knowledge is declared as relevant.

Regarding transversal skills (see Figure 4B), the pattern between social skills and the rest of the skills is clearly different. While the latter are generally associated with a lower risk of overeducation, social skills show non-significant or even increasing effects, such as for health sciences graduates. Although the effect is heterogenous across fields, IT, management and language skills are associated, overall, with a lower risk of overeducation. Despite the effect being heterogenous in magnitude, management and language skills significantly lower overeducation in almost all fields. The only exception is observed among health sciences graduates, for whom language skills increase the risk of overeducation, although the effect is not statistically significant. For IT skills, the reduction in the risk of overeducation is only statistically significant among engineering and architecture graduates.

These overeducation effects of knowledge and skills might mask interesting and possibly heterogeneous patterns in the interconnection between both. In order to shed more light on this heterogeneity, Table 4 shows the role of these skills in the probability of overeducation depending on whether theoretical and/or practical knowledge have been important to get the current job. Overall, our findings show that management and language skills significantly reduce the risk of overeducation regardless of the relevance of theoretical or practical knowledge. For IT skills, this reduction is only observed when theoretical knowledge was not relevant for getting the current job.

---

[12] The findings presented in this subsection correspond to a new specification that includes interactions between theoretical/practical knowledge, transversal skills and field of study. These interactions would lead to many cells with very few observations. Thus, we considered binary indicators for both theoretical/practical knowledge and transversal skills. These binary indicators take the value of 1 if a given skill or knowledge has been important or very important to get the job and 0 otherwise.

**Figure 4:**
**Relevance of knowledge and skills on the probability of overeducation in the current job**
(Average partial effects in percentage points)

**4A: Theoretical and practical knowledge**

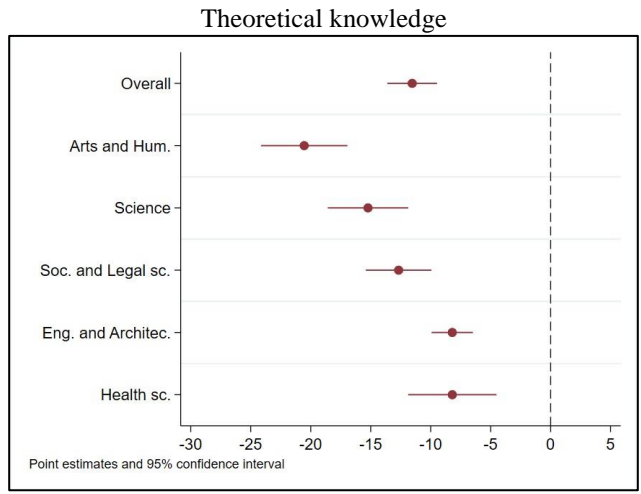
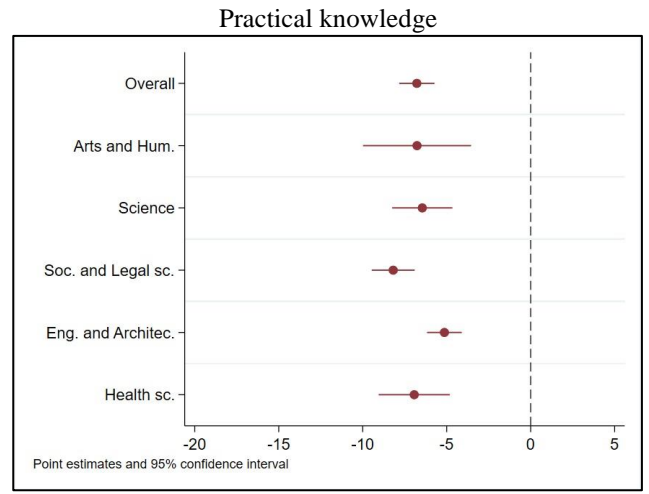

**4B: Transversal skills**

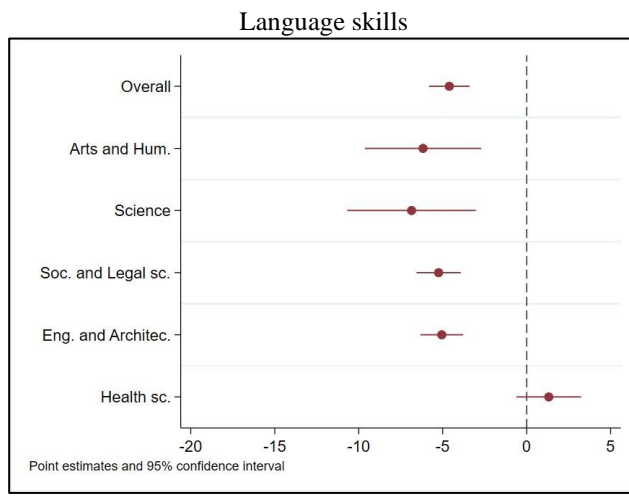
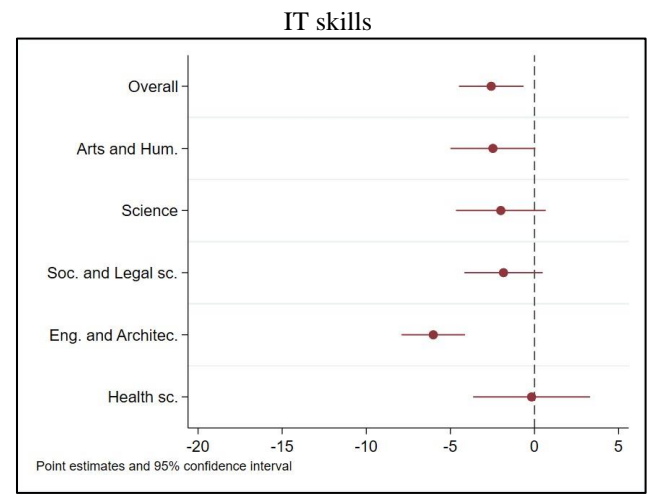

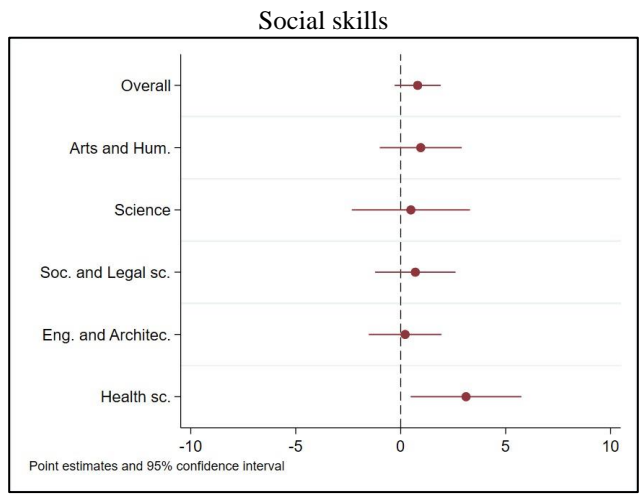
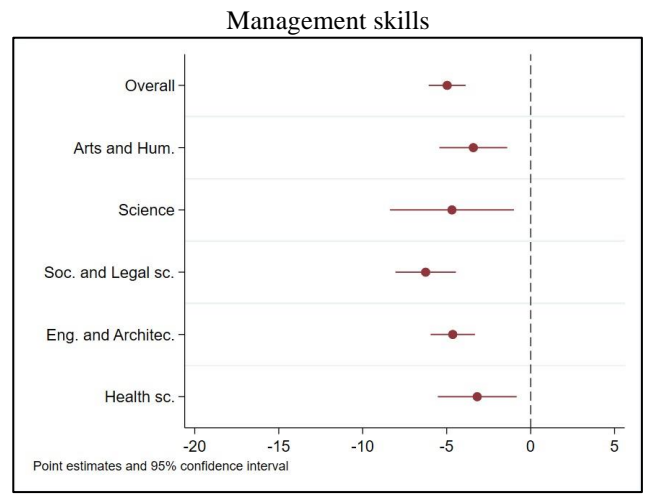

As regards the magnitude of the role of the different skills, it is remarkable that for language, management and IT skills, the reduction is higher when theoretical knowledge was not relevant to obtain the current job, regardless of the importance of practical knowledge. This suggests some kind of trade-off between theoretical knowledge and skills. For instance, graduates who claim that language skills have been important to get their job, exhibit – overall and regardless of the relevance of practical knowledge – an almost 7 pp reduction in their risk of overeducation when their theoretical knowledge is declared as non-relevant. The corresponding figure is halved when this type of knowledge is relevant. A very different pattern is found for social skills. Overall, we obtain that having declared social skills as important to obtain the job while not having achieved a match in terms of theoretical and practical knowledge significantly increases the risk of overeducation. However, once either practical or theoretical knowledge are declared as necessary to get the job, the effect of social skills becomes non-significant.

When examining the interconnection between skills and knowledge across fields of study, the most striking findings appear for engineering and architecture and health sciences graduates. For the former, IT skills are revealed as important competencies to significantly reduce the risk of overeducation, regardless of the relevance of theoretical or practical knowledge. Still, the abovementioned compensation effect seems to be in place since the magnitude of this reduction is clearly higher when theoretical knowledge is non-relevant. Concerning health sciences graduates, we only obtain evidence of this compensation effect for management skills. Moreover, it seems that declaring social skills as relevant may increase the probability of overeducation among these graduates, especially when practical knowledge is declared as non-relevant. The latter might be explained by the type of job performed by well-matched health sciences graduates for whom the most important skills are possibly those acquired during their degree programmes and are therefore field-specific skills that cannot be easily obtained otherwise.

Our results thus confirm the overall importance of the interconnection between knowledge and skills to determine graduates' career success in the transition process from school to work. This suggests the need to identify overarching goals for education systems and lifelong learning in order to ensure that young adults are adequately prepared for todays' labour market demands. In particular, all programmes of study should give students the chance to connect academic knowledge with the skills needed to be successful in their professional lives. This need to combine knowledge with skills should not only concern policymakers, but also educators insofar they are responsible for finding the way to empower, motivate and engage their students by redefining their methodologies and curricular plans accordingly (Greenberg and Nilssen, 2015). All in all, our results suggest that traditional education systems should adopt a more "competence-based" approach (Hall and Jones, 1976; Riesman, 1979; Johnstone and Soares, 2014) that enables the construction of well-qualified professionals endowed with transversal skills that will enable them to adapt to the ever-changing labour markets.

**Table 4. Importance of knowledge and transversal skills (average partial effects on the probability of overeducation)**

| | Theoretical knowledge (NR) | | Theoretical knowledge (R) | |
|---|---|---|---|---|
| | Practical knowledge (NR) | Practical knowledge (R) | Practical knowledge (NR) | Practical knowledge (R) |
| **Language skills** | | | | |
| Overall | -0.068***(0.017) | -0.066***(0.008) | -0.037***(0.013) | -0.033***(0.004) |
| Arts and humanities | -0.102***(0.033) | -0.102***(0.026) | -0.047**(0.022) | -0.041***(0.013) |
| Science | -0.116***(0.038) | -0.104***(0.024) | -0.056***(0.022) | -0.044***(0.012) |
| Social and legal sciences | -0.083***(0.019) | -0.077***(0.009) | -0.044***(0.014) | -0.037***(0.004) |
| Engineering and architecture | -0.075***(0.012) | -0.068***(0.010) | -0.044***(0.011) | -0.037***(0.007) |
| Health Sciences | 0.025(0.023) | 0.008(0.012) | 0.023*(0.014) | 0.009*(0.005) |
| **IT skills** | | | | |
| Overall | -0.068***(0.023) | -0.051***(0.016) | -0.001(0.013) | 0.001(0.007) |
| Arts and humanities | -0.076***(0.030) | -0.061***(0.018) | 0.004(0.018) | 0.005(0.009) |
| Science | -0.063**(0.027) | -0.048*(0.025) | 0.006(0.011) | 0.006(0.010) |
| Social and legal sciences | -0.064**(0.026) | -0.047***(0.017) | 0.009(0.017) | 0.008(0.009) |
| Engineering and architecture | -0.116***(0.021) | -0.088***(0.017) | -0.041***(0.011) | -0.028***(0.006) |
| Health sciences | -0.025(0.034) | -0.013(0.023) | 0.021(0.018) | 0.013(0.010) |
| **Social skills** | | | | |
| Overall | 0.040***(0.012) | 0.005(0.008) | 0.012(0.017) | -0.008(0.011) |
| Arts and humanities | 0.048**(0.021) | 0.008(0.017) | 0.013(0.018) | -0.008(0.013) |
| Science | 0.035(0.032) | 0.001(0.019) | 0.007(0.021) | -0.010(0.011) |
| Social and legal sciences | 0.041**(0.019) | 0.004(0.010) | 0.012(0.022) | -0.010(0.014) |
| Engineering and architecture | 0.025***(0.009) | -0.002(0.011) | 0.005(0.017) | -0.011(0.017) |
| Health sciences | 0.073***(0.021) | 0.030(0.020) | 0.032**(0.014) | 0.009(0.010) |
| **Management skills** | | | | |
| Overall | -0.093***(0.012) | -0.075***(0.005) | -0.037***(0.012) | -0.026***(0.006) |
| Arts and humanities | -0.075***(0.020) | -0.066***(0.014) | -0.016(0.015) | -0.012(0.008) |
| Science | -0.094***(0.035) | -0.078***(0.025) | -0.030(0.023) | -0.021(0.013) |
| Social and legal sciences | -0.117***(0.017) | -0.095***(0.010) | -0.050***(0.017) | -0.035***(0.008) |
| Engineering and architecture | -0.083***(0.011) | -0.067***(0.009) | -0.034***(0.009) | -0.026***(0.006) |
| Health sciences | -0.070***(0.023) | -0.046***(0.013) | -0.019(0.015) | -0.011(0.007) |

*Notes:* Average partial effects in percentage points.
NR: The corresponding knowledge or skill has been declared as non-relevant for the graduate to attain the current job.
R: The corresponding knowledge or skill has been declared as relevant to for the graduate to attain the current job.
Results from a trivariate probit model estimation. Clustered standard errors by occupation obtained by the Delta method are shown in parentheses. These results have been obtained by adding to the baseline model in section 4 interaction terms between the field of study and knowledge and the skills variables, as well as interaction terms between knowledge and the skills variables.
***, **, *: statistically significant at 1%, 5% and 10%, respectively.

## 5 Conclusions

In this paper we have studied the incidence and persistence of overeducation among university graduates in the Spanish labour market. The aim of our analysis has been to understand how both indicators behave depending on the economic scenario that graduates face when entering the labour market. We have also studied how the incidence and persistence of overeducation across fields of study might have evolved under different economic settings, which, to the best of our knowledge, has not been previously explored in the literature. Also, as a novelty, we have explored the potential interconnection between transversal skills and practical or theoretical knowledge in determining the risk of overeducation, again, across fields of study.

For the analysis we have used the Survey on the Labour Insertion of University Graduates (EILU) in Spain conducted by the National Statistics Institute (INE) in 2014 and 2019. The EILU offers information on the early career of graduates in 2010 and in 2014 under an economic downturn and an economic expansion, respectively. The dataset is rich in individual and job characteristics and provides information on the first job and the job five years after graduation, thus allowing different sources of endogeneity to be considered in the analysis.

Our results show, first, that graduating in a period of recession significantly increases the risk of overeducation five years later. Second, overeducation is a persistent phenomenon that is also shaped by the economic cycle: regardless of the economic scenario at the time of graduation, the risk of overeducation five years later is higher for those overeducated in the first job, and the persistence is more pronounced for those who graduated in an economic downturn. Third, there is evidence of heterogeneity across fields of study regarding the dependence of the risk and persistence of overeducation on the economic cycle. Overall, health sciences graduates exhibit the lowest incidence and persistence of overeducation and are less affected by the business cycle in this regard. In contrast, arts and humanities graduates suffer the highest incidence and persistence of overeducation, regardless of the economic period. Social sciences, science and, especially, engineering and architecture graduates' overeducation risk and persistence seem to be more dependent on the economic cycle.

Finally, we find evidence of potential interconnections between knowledge inherent to the different fields of study and other types of transversal skills in determining graduates' success in the transition process from school to work. In particular, some kind of transversal skills (language, management and IT skills) seem to be especially effective in reducing the risk of overeducation when the graduates did not achieve an appropriate job match in terms of their theoretical knowledge.

Overall, although investing in tertiary education is an essential tool for securing employment, employability and quality of employment also depend on having specific knowledge and skills sets that enable individuals not only to get the appropriate jobs but to remain in employment and advance in their careers. In particular, our findings reveal that overeducation continues to be a critical issue among recent university graduates that is highly conditioned by the knowledge and skills set needed for their jobs. Insofar as this job education mismatch implies a loss in potential productivity, coping with this problem will require the better development of individuals' abilities to tackle complex mental tasks, going well beyond the basic reproduction of accumulated knowledge.

Policymakers should promote measures that, beyond increasing the percentage of university graduates, aim to equip them with the necessary skills to meet the demands of an ever-changing labour market. Looking at our results and considering the differences across fields of study, those policies could be especially important among graduates from fields that exhibit a higher risk of unemployment and overeducation. Governments should invest in information campaigns to better guide former university students about the positions demanded in the labour market, as well as the risk of overeducation and persistence associated with each field of study. This could potentially lead to more efficient educational choices from the labour supply side and avoid job-education mismatches in the future. Moreover, insofar as such demands are highly shaped by the business cycle, effective measures intended to reduce job-education mismatches and to avoid a potential crowding-out effect of the less educated should take macroeconomic conditions into consideration. In addition, it is important to note that overeducation is not only a supply-side problem and policies that favour the creation of job positions demanding high-educated workers should therefore be contemplated. Other policies that help graduates improve their skills to find a suitable job and overcome difficulties such as geographic mobility would also foster an equilibrium between the supply and demand of highly educated individuals.

Important steps have already been taken in the global context to address the problem of education and skill mismatches. For instance, the 2030 European strategic framework for education and training has highlighted the need to prevent skills gaps and education mismatches and actions such as those included in the European Skills Agenda have this aim. An important concern in this context is the development of an overarching conceptual framework based on broad theories of what skills, knowledge and competencies are and how they relate to each other. The OECD Programme Definition and Selection of Competencies: Theoretical and Conceptual Foundations (DeSeCo) was initiated to work towards filling this gap. Initiatives such these should become a major priority for the development and maintenance of human and social capital, which represent important factors for societies not only to achieve prosperity and social cohesion, but first and foremost manage the challenges and tensions of an increasingly interdependent and changing world.

Dolado, J.J., Felgueroso, F., & Jimeno, J.F. (2000). Youth labour markets in Spain: Education, training, and crowding-out. *European Economic Review, 44*(4–6), 943–956.

Dolton, P., & Vignoles, A. (2000). The incidence and effects of overeducation in the UK graduate labour market. *Economics of Education Review*, *19*(2), 179–198.

Duncan, G.J., & Hoffman, S.D. (1981). The incidence and wage effects of overeducation. *Economics of education review*, *1*(1), 75–86.

Eby, L.T., Butts, M., & Lockwood, A. (2003). Predictors of success in the era of the boundaryless career. *Journal of Organizational Behavior, 24*, 689–708.

Eckaus, R.S. (1964). Economic criteria for education and training. *The Review of Economics and Statistics*, *47*(2), 181–190.

Erdsiek, D. (2016). Overqualification of graduates: assessing the role of family background. *Journal for Labour Market Research*, *49*(3), 253–268.

Erdsiek, D. (2021). Dynamics of overqualification: evidence from the early career of graduates. *Education Economics, 29*(3), 312–340.

Ermini, B., Papi, L., & Scaturro, F. (2017). An analysis of the determinants of over-education among Italian Ph. D graduates. *Italian Economic Journal*, *3*, 167–207.

Flisi, S., Goglio, V., Meroni, E., Rodrigues, M., & Vera-Toscano, E. (2017). Measuring Occupational Mismatch: Overeducation and Overskill in Europe-Evidence from PIAAC. *Social Indicators Research, 131*(3), 1211–1249.

Freeman, R. (1976). *The overeducated American*. Academic Press

Frei, C., & Sousa-Poza, A. (2012). Overqualification: permanent or transitory? *Applied Economics*, *44*(14), 1837–1847.

Frenette, M. (2004). The overqualified Canadian graduate: the role of the academic programme in the incidence, persistence, and economic returns to overqualification. *Economics of Education Review*, *23*(1), 29–45.

Garcia-Mainar, I., & Montuenga, V.M. (2019). The signalling role of over-education and qualifications mismatch. *Journal of Policy Modeling, 41*(1), 99-119.

García-Montalvo, J. (1995). Empleo y sobrecualificación: el caso español. *Documento de trabajo, (95–20)*, FEDEA.

García-Montalvo, J. (2009). La inserción laboral de los universitarios y el fenómeno de la sobrecualificación en España. *Papeles de economía española*, *119*, 172–187.
30

# Appendix

**Table A1. Knowledge and skills: summary of descriptive statistics**

How important has this knowledge/skill been to obtain the current job? Binary indicators for each category. N = 37,819 observations.

| Variable | Mean | Std. Dev. |
|---|---|---|
| Theoretical knowledge | | |
|     Not important | 0.106 | 0.308 |
|     Slightly important | 0.114 | 0.318 |
|     Moderately important | 0.183 | 0.387 |
|     Important | 0.296 | 0.457 |
|     Very important | 0.300 | 0.458 |
| Practical knowledge | | |
|     Not important | 0.091 | 0.287 |
|     Slightly important | 0.084 | 0.278 |
|     Moderately important | 0.139 | 0.346 |
|     Important | 0.283 | 0.451 |
|     Very important | 0.403 | 0.491 |
| Language skills | | |
|     Not important | 0.247 | 0.431 |
|     Slightly important | 0.181 | 0.385 |
|     Moderately important | 0.178 | 0.382 |
|     Important | 0.170 | 0.375 |
|     Very important | 0.225 | 0.417 |
| IT skills | | |
|     Not important | 0.136 | 0.342 |
|     Slightly important | 0.144 | 0.351 |
|     Moderately important | 0.229 | 0.420 |
|     Important | 0.286 | 0.452 |
|     Very important | 0.206 | 0.404 |
| Social skills | | |
|     Not important | 0.052 | 0.222 |
|     Slightly important | 0.042 | 0.200 |
|     Moderately important | 0.105 | 0.306 |
|     Important | 0.314 | 0.464 |
|     Very important | 0.487 | 0.500 |
| Management skills | | |
|     Not important | 0.070 | 0.255 |
|     Slightly important | 0.071 | 0.257 |
|     Moderately important | 0.153 | 0.360 |
|     Important | 0.334 | 0.472 |
|     Very important | 0.372 | 0.483 |

**Table A2. Estimated average partial effects on overeducation in current job**
**(Trivariate probit model)**

| | | | |
|---|---|---|---|
| Recovery period | -0.106***(0.018) | Overeducation in the first job | 0.201***(0.013) |
| **Individual characteristics** | | **Job-related variables** | |
| Male | -0.004(0.004) | Part-time job | 0.054***(0.009) |
| Spanish | 0.011 (0.017) | Professional sit. (ref: Trainee) | |
| Age intervals (ref: <30 years) | |     Permanent contract | 0.037***(0.008) |
|     30–34 years old | 0.008(0.006) |     Fixed-term contract | 0.039***(0.009) |
|     >34 years old | 0.025***(0.007) | More than two employers | -0.020 (0.026) |
| IT knowledge (ref: Basic) | | More than two years of | -0.030***(0.009) |
|     Advanced | 0.000(0.004) | Theoretical knowledge (ref: | |
|     Expert | -0.012*(0.007) |     Moderately important | -0.139***(0.011) |
| More than one language spoken | -0.003(0.005) |     Very important | -0.212***(0.014) |
| | | Practical knowledge (ref: None) | |
| **Study-related variables** | |     Moderately important | -0.037***(0.008) |
| Studied abroad | -0.015***(0.005) |     Very important | -0.085***(0.007) |
| Coll. or excellence grant | -0.014***(0.004) | Languages skills (ref: None) | |
| Private university | -0.021***(0.003) |     Moderately important | -0.016***(0.005) |
| Field of study (Ref: Arts and humanities) | |     Very important | -0.066***(0.007) |
|     Science | -0.035***(0.005) | IT skills (ref: None) | |
|     Social and legal sciences | -0.019***(0.005) |     Moderately important | -0.036***(0.009) |
|     Eng. and arch. | -0.060***(0.008) |     Very important | -0.032**(0.014) |
|     Health sciences | -0.109***(0.010) | Soc skills (ref: None) | |
| Internship outside degree | -0.004(0.003) |     Moderately important | 0.050***(0.013) |
| Postgraduate studies | -0.027***(0.005) |     Very important | 0.056***(0.017) |
| | | Management skills (ref: None) | |
| | |     Moderately important | -0.030***(0.011) |
| | |     Very important | -0.073***(0.013) |
| *Observations* | 37819 | | |
| corr($u_f, u_E$) | 0.354***(0.057) | | |
| corr($u_f, u_c$) | 0.283***(0.024) | | |
| corr($u_E, u_c$) | -0.066(0.050) | | |
| *p-value (Joint significance tests)* | | | |
| All variables | 0.000 | | |
| Individual characteristics | 0.000 | | |
| Study variables | 0.000 | | |
| Job-related variables | 0.000 | | |
| Regional (or countries) dummies | 0.000 | | |
| Interaction terms | 0.000 | | |

*Notes*: Average partial effects from a trivariate probit model estimation. Clustered standard errors by occupation obtained by the Delta method are shown in parentheses. Regional dummies and intermediate skills categories included in the model, but results are not reported due to limitations of space. They are available upon request.
\*\*\*, \*\*, \*: statistically significant at 1%, 5% and 10%, respectively.

**Table A3. Estimated APE on the probability of overeducation in the job at the time of the interview (Bivariate probit model)**

|  | Full sample | 2014 wave (Graduated in 2010) | 2019 wave (Graduated in 2014) |
|---|---|---|---|
| Recovery period | -0.103***(0.018) | | |
| Overeducation in the first job | 0.186***(0.010) | 0.282***(0.018) | 0.117****(0.008) |
| Field of study (Ref: Arts and humanities) | | | |
|    Science | -0.033***(0.005) | -0.021***(0.005) | -0.043***(0.007) |
|    Social and legal sciences | -0.017**(0.005) | -0.003(0.007) | -0.028***(0.006) |
|    Engineering and architecture | -0.057***(0.008) | -0.040***(0.011) | -0.072***(0.012) |
|    Health sciences | -0.105***(0.010) | -0.122***(0.015) | -0.093***(0.010) |
| Observations | 37,819 | 37,819 | 37,819 |
| Wald test [p-value] | 0.000 | | |
| corr($u_f, u_c$) | 0.306***(0.024) | | |

*Notes:* APE: Average partial effects from a bivariate probit model estimation. Clustered standard errors by occupation obtained by the Delta method are shown in parentheses. Individual characteristics, study-related variables, job-related variables, interaction terms (see Section 3.2) and regional dummies included in all models.
***, **, *: statistically significant at 1%, 5% and 10%, respectively.